\documentclass[fleqn,usenatbib]{mnras}

\usepackage{newtxtext,newtxmath}

\usepackage[T1]{fontenc}

\DeclareRobustCommand{\VAN}[3]{#2}
\let\VANthebibliography\thebibliography
\def\thebibliography{\DeclareRobustCommand{\VAN}[3]{##3}\VANthebibliography}

\usepackage{graphicx}	
\usepackage{amsmath}	

\usepackage{amssymb}	
\usepackage{natbib}

\usepackage{subcaption}
\usepackage{gensymb}

\newcommand{\HI}{{\sc H\,i }}

\title[Dark matter in intermediate-mass ETGs]{Dark matter measurements combining stellar and \HI kinematics: 30\% $1-\sigma$ outliers with low dark matter content at $5R_\mathrm{e}$}

\author[Yang et al.]{Meng Yang$^{1}$\thanks{E-mail: myang@shao.ac.cn},
Ling Zhu$^{1}$\thanks{E-mail: lzhu@shao.ac.cn},
Yu Lei$^{1}$,
Nicholas Boardman$^{2}$, 
Anne-Marie Weijmans$^{2}$, \newauthor
Raffaella Morganti$^{3,4}$,
Tom Oosterloo$^{3,4}$,
Pierre-Alain Duc$^{5}$,
\\
$^{1}$Shanghai Astronomical Observatory, Chinese Academy of Sciences, 80 Nandan Road, Shanghai 200030, China\\
$^{2}$School of Physics and Astronomy, University of St Andrews, North Haugh, St Andrews, KY16 9SS, UK\\
$^{3}$Netherlands Institute for Radio Astronomy (ASTRON), Postbus 2, NL-7990 AA Dwingeloo, the Netherlands\\
$^{4}$Kapteyn Astronomical Institute, University of Groningen, PO Box 800, NL-9700 AV Groningen, the Netherlands\\
$^{5}$Observatoire astronomique de Strasbourg, 11 rue de l'université, 67000 Strasbourg, France
}

\date{Accepted XXX. Received YYY; in original form ZZZ}

\pubyear{2024}

\begin{document}
\label{firstpage}
\pagerange{\pageref{firstpage}--\pageref{lastpage}}
\maketitle

\begin{abstract}
We construct the Schwarzschild dynamical models for 11 early-type galaxies with the SAURON and Mitchell stellar IFUs out to $2-4 R_\mathrm{e}$, and construct dynamical models with combined stellar and \HI kinematics for a subsample of 4 galaxies with \HI velocity fields out to $10 R_\mathrm{e}$ obtained from the Westerbork Synthesis Radio Telescope, thus robustly obtaining the dark matter content out to large radii for these galaxies. Adopting a generalised-NFW dark matter profile, we measure an NFW-like density cusp in the dark matter inner slopes for all sample galaxies, with a mean value of $1.00\pm0.04$ (rms scatter $0.15$). The mean dark matter fraction for the sample is $0.2$ within $1 R_\mathrm{e}$, and increases to $0.4$ at $2 R_\mathrm{e}$, and $0.6$ at $5 R_\mathrm{e}$. The dark matter fractions within $1 R_\mathrm{e}$ of these galaxies are systematically lower than the predictions of both the TNG-100 and EAGLE simulations. For the dark matter fractions within $2 R_\mathrm{e}$ and $5 R_\mathrm{e}$, 40\% and 70\% galaxies are $1-\sigma$ consistent with either the TNG-100 or the EAGLE predictions, while the remaining 60\% and 30\% galaxies lie below the $1-\sigma$ region. Combined with 36 galaxies with dark matter fractions measured out to $5 R_\mathrm{e}$ in the literature, about 10\% of these 47 galaxies lie below the $3-\sigma$ region of the TNG-100 or EAGLE predictions.
\end{abstract}





\section{Introduction}


In the context of cold dark matter cosmology~\citep{1984Natur.311..517B,1985ApJ...292..371D}, galaxies form in the centres of dark matter halos. The distribution of dark matter within galaxies is affected not only by the characteristics of dark matter particles~\citep[e.g.][]{2016PhRvL.116d1302K,2018MNRAS.476L..20R} but also by the presence of baryonic matter during galaxy formation. Star formation is regulated by a series of feedback mechanisms, such as active galactic nucleus feedback~\citep[e.g.][]{2005Natur.433..604D,2006ApJS..163....1H,2010MNRAS.406L..55D} and supernova feedback~\citep[e.g.][]{2009ApJ...697.2030S,2011ApJ...731...41O}. These processes transfer energy to dark matter particles through dynamical friction, thus altering the distribution of dark matter within galaxies~\citep[e.g.][]{1996MNRAS.283L..72N,2005MNRAS.356..107R,2011MNRAS.416.1118C}. In shallow gravitational potential wells, such as those of dwarf galaxies, baryonic feedback has a noticeable effect on the dark matter distribution, which can explain the observed "cored" dark matter distribution~\citep{2019MNRAS.484.1401R}. However, for massive galaxies, baryonic feedback might not be enough to significantly modify the distribution of dark matter~\citep{2016MNRAS.456.3542T,2018MNRAS.480..800H,2017ARA&A..55..343B}. 

Various techniques are employed to investigate dark matter in galaxies, such as strong lensing~\citep[e.g.][]{2008ApJ...684..248B,2012MNRAS.423.1073B,2014MNRAS.439.2494O,2022MNRAS.510L..24L}, hot interstellar X-ray gas~\citep[e.g.][]{2006ApJ...636..698F,2019ApJ...887..259H,2020ApJ...905...28H} and dynamical methods. Rotation curves have been widely used to late-type galaxies in the local universe~\citep[e.g.][]{2008AJ....136.2761O,2015AJ....149..180O,2013A&A...557A.131M,2019ApJ...887...94R,2019MNRAS.489.5483T} and at higher redshift $z \sim 1-2$~\citep[e.g.][]{2019MNRAS.485..934T,2020ApJ...902...98G,2021A&A...653A..20S,2022A&A...659A..40S,2023ApJ...944...78N}, but their applications in early-type galaxies are limited. Dynamical modelling is a key method for accurately determining the dark matter distributions in early-type galaxies with complicated dynamical structures: spherical Jeans models~\citep{1982MNRAS.200..361B,1983ApJ...266...58T} suffer from a mass-anisotropy degeneracy, which requires further assumptions to infer the mass distribution; anisotropic Jeans models~\citep{2008MNRAS.390...71C} reduce ad-hoc assumptions of underlining structures for axisymmetric galaxies; particle-based made-to-measure models~\citep{1996MNRAS.282..223S} and orbit-based Schwarzschild models~\citep{1979ApJ...232..236S} sidestep this degeneracy and succeed in modelling triaxial galaxies. 

Most studies that assess the dark matter mass fraction of large samples with dynamical modelling rely on stellar kinematic data from spatially resolved integral field unit (IFU) spectrographs, such as Atlas$\rm ^{3D}$~\citep{2011MNRAS.413..813C}, CALIFA~\citep{2012A&A...538A...8S}, SAMI~\citep{2012MNRAS.421..872C}, and SDSS-IV MaNGA~\citep{2015ApJ...798....7B}. However, these data only cover the brightest central 1-2 effective radii ($R_\mathrm{e}$) of galaxies, where the contribution of dark matter is minimal. Therefore, using only these stellar kinematic data to measure the properties of dark matter halos (including the dark matter fraction and density profile) can lead to considerable uncertainties~\citep{2013MNRAS.432.1709C,2018MNRAS.473.3000Z,2022ApJ...930..153S,2023MNRAS.522.6326Z}.

In order to obtain more precise measurements of dark matter, we require kinematic data extending to the outskirts of galaxies. Although stellar kinematics using extended IFU spectrographs and long-slit spectrographs~\citep{2009MNRAS.398..561W,2010ApJ...716..370F,2015ApJ...804L..21C,2016MNRAS.460.3029B,2023A&A...675A.143C} can provide such measurements, the amount of data is limited by the required long exposure time. Therefore, other extended tracers are often used to measure dark matter distributions. Planetary nebulae (PNe) and globular clusters (GCs) are commonly employed for dark matter measurements independently~\citep{2003Sci...301.1696R,2001ApJ...559..828C,2003ApJ...591..850C,2012ApJ...748....2D,2016MNRAS.460.3838A,2017MNRAS.468.3949A,2023arXiv230611786D}, or in combination with stellar kinematics~\citep{2008MNRAS.385.1729D,2009MNRAS.395...76D,2009MNRAS.393..329N,2011MNRAS.411.2035N,2014MNRAS.439..659N,2011MNRAS.415.1244D,2013MNRAS.431.3570M,2014ApJ...792...59Z,2016MNRAS.462.4001Z,2020MNRAS.492.2775L,2022RAA....22h5023Y}, mostly for the most massive early-type galaxies, where there are enough of these tracers to make measurements. On the other hand, the \HI gas rotation curve is a powerful tracer of dark matter mass in late- and early-type galaxies \citep{2008AJ....136.2648D,2011AJ....141..193O,2016AJ....152..157L}. Combining \HI gaseous kinematics and stellar IFUs can provide strong constraints on the dark matter fraction of galaxies from the inner to outer regions~\citep{2020MNRAS.491.4221Y}.

Current measurements of dark matter fractions in galaxies, in comparison to cosmological numerical simulations~\citep{2018MNRAS.481.1950L}, demonstrate the following: (1) Different cosmological numerical simulations do indeed produce varying dark matter fractions within $5 R_\mathrm{e}$; (2) Observations that reach out to $5 R_\mathrm{e}$ are inadequate and display a considerable scatter, with few measurements in galaxies with stellar masses below $10^{10} M_\odot$; (3) Dark matter fractions measured using stellar kinematics only covering $1-2 R_\mathrm{e}$ may be affected by systematic biases due to the degeneracy between dark matter and baryonic matter. Consequently, the existing observational data are not enough to impose stringent constraints on cosmological numerical simulations. To effectively constrain cosmological numerical simulations, precise measurements of dark matter fractions are needed from large samples extending to large radii, while breaking the degeneracy between dark matter and baryonic matter.


This paper aims to measure the dark matter profiles of a selection of intermediate mass early-type galaxies with extended kinematics. We structure this paper as follows: Section~\ref{sec:sample} introduces the sample and data used in our dynamical models. Section~\ref{sec:method} outlines our dynamical modelling method. Section~\ref{sec:result} presents our results, including the dark matter fractions and the total density slopes. In Section~\ref{sec:discussion}, we discuss our results and compare them with simulations and observations in the literature. Finally, Section~\ref{sec:summary} summarises our work.

\section{Sample and data}
\label{sec:sample}
\begin{figure}
    \centering
    \includegraphics[width=1.0\columnwidth]{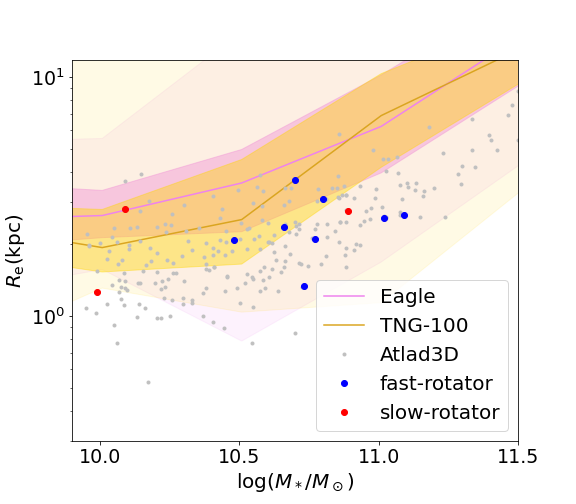}
    \caption{Mass-size relation of all sample galaxies, with the fast-rotators marked in blue and slow-rotators marked in red. The $M_*$ here is calculated from $M_K$ according to Equation 2 of~\citet{2013MNRAS.432.1709C}. The grey dots are all the galaxies in the Atlas$\rm ^{3D}$ Survey~\citep{2011MNRAS.413..813C}. The gold and violet solid curves are the mass-size relations of the early-type galaxies in the TNG-100 and EAGLE simulations, and the corresponding dark and light shaded areas cover the 16th to 84th fractiles and 0.15th to 99.85th fractiles, respectively.} 
    \label{fig:mass-size}
\end{figure}

We choose the sample of 12 ETGs presented in~\citep{2017MNRAS.471.4005B}, with the exception of NGC 3626 due to its strong bar~\citep{2015ApJS..219....4S}. Details of the 11 galaxies in our sample are shown in Table~\ref{tab:sample-info}. This sample has a stellar mass range of $10^{10}-10^{11} M_\odot$ and a range in effective radius of $1-4$ kpc, and includes 3 slow-rotators dominated by random motions and 8 fast-rotators dominated by circular motions, as classified in~\citet{2011MNRAS.413..813C}. We show the mass-size relation of the sample in Figure~\ref{fig:mass-size}, and compare it with the full Atlas$\rm ^{3D}$ sample. We also plot the mass-size relations of the early-type galaxies in the TNG-100 simulation~\citep{2018MNRAS.477.1206N,2018MNRAS.480.5113M,2018MNRAS.475..648P,2018MNRAS.475..624N,2018MNRAS.475..676S} and the EAGLE simulation~\citep{2015MNRAS.446..521S,2015MNRAS.450.1937C}, in which the $R_\mathrm{e}$ of a simulated galaxy is obtained by projecting this galaxy along a random direction to obtain its surface brightness~\citep{2018MNRAS.474.3976G}, and the early-type galaxies are defined as central galaxies with a specific star formation rate (sSFR) lower than $10^{-11}/\mathrm{yr}$ in both simulations. Cosmological simulations have not been able to accurately replicate the mass-size relation seen in observations, especially at the high-mass end~\citep{2018MNRAS.481.1950L}, which could lead to discrepancies when using $R_\mathrm{e}$ to measure enclosed dark matter fractions.
                  
\begin{table*}
    \renewcommand\arraystretch{1.2}
    \centering
    \caption{Basic information of sample galaxies. All values are taken from~\citet{2011MNRAS.413..813C}, except that the K-band stellar mass $M_{*,K}$ is estimated from $\mathrm{M}_K$ according to Equation 2 in~\citet{2013ApJ...778L...2C}, and $R_\mathrm{e}$ in kpc is calculated from $R_\mathrm{e}$ in arcsec and distance. FR / SR stands for fast-rotators / slow-rotators as classified in~\citet{2011MNRAS.413..813C}.}
    \begin{tabular}{ccccccccc}
    \hline
    Galaxy & RA(\degree) & DEC(\degree) & FR / SR & Distance (Mpc) & $\mathrm{M}_K$ & $ \mathrm{log} (M_{*,K}/M_\odot)$  & $R_\mathrm{e}$ (arcsec) & $R_\mathrm{e}$ (kpc)  \\\hline
    NGC0680 & 27.447035  & 21.970827 & FR & 37.5 & -24.17 & 11.09 & 14.5  & 2.64 \\
    NGC1023 & 40.100052  & 39.063251 & FR & 11.1 & -24.01 & 11.02 & 47.9  & 2.58 \\
    NGC2685 & 133.894791 & 58.734409 & FR & 16.7 & -22.78 & 10.48 & 25.7  & 2.08 \\
    NGC2764 & 137.072983 & 21.443447 & FR & 39.6 & -23.19 & 10.66 & 12.3  & 2.36 \\
    NGC3522 & 166.668549 & 20.085621 & SR & 25.5 & -21.67 & 9.99  & 10.2  & 1.26 \\
    NGC3998 & 179.484039 & 55.453564 & FR & 13.7 & -23.33 & 10.73 & 20.0  & 1.33 \\
    NGC4203 & 183.770935 & 33.197243 & FR & 14.7 & -23.44 & 10.77 & 29.5  & 2.1 \\
    NGC5582 & 215.179703 & 39.693584 & FR & 27.7 & -23.28 & 10.7  & 27.5  & 3.69 \\
    NGC5631 & 216.638687 & 56.582664 & SR & 27.0 & -23.70 & 10.89 & 20.9  & 2.74 \\
    NGC6798 & 291.013306 & 53.624752 & FR & 37.5 & -23.52 & 10.8  & 16.9  & 3.07 \\
    UGC03960 & 115.094856 & 23.275089 & SR & 33.2 & -21.89 & 10.09 & 17.4  & 2.8 \\
    \hline
    \end{tabular}
    \label{tab:sample-info}
\end{table*}

We used the $r$-band images obtained with the MegaCam instrument on the Canada-France-Hawaii Telescope (CFHT) as part of the MATLAS survey~\citep{2015MNRAS.446..120D}. This instrument has a wide field-of-view of $1$ degree$^2$, and these images have a limiting magnitude of $\sim 29$ mag arcsec$^{-2}$ in the $r$-band, which allows us to trace the surface brightness out to $5 R_\mathrm{e}$, beyond the range of our kinematic data sets. As the MATLAS $r$-band images of most sample galaxies (all except NGC 2764, NGC 6798 and UGC 03960) are saturated in the galaxy centre, we also use their $r$-band images taken from the Panoramic Survey Telescope and Rapid Response System (Pan-STARRS) for the correction of saturation, which are stacked from short-exposure images to reach a limiting magnitude of $23.2$ in the $r$-band. We refer the reader to~\citet{2016arXiv161205560C} and the references therein for further information on Pan-STARRS. 

This sample of galaxies has two sets of stellar kinematic data. 
The central stellar kinematics were observed with the SAURON integral field spectrograph on the William Herschel Telescope~\citep{2001MNRAS.326...23B} in the Atlas$\rm ^{3D}$ Survey~\citep{2011MNRAS.413..813C}. The SAURON integral field spectrograph has a field-of-view of $33\times 41$ arcsec$^2$ with lenslets of $0.94$ arcsec$^2$ and an instrumental velocity dispersion of $105$ km/s, and covers the central $1 R_\mathrm{e}$ of the sample galaxies. The spectra were Voronoi binned~\citep{2003MNRAS.342..345C} to a signal-to-noise ratio of 40, and fitted with the penalised pixel fitting method~\citep[pPXF;][]{2004PASP..116..138C} to obtain the stellar kinematics (velocity, velocity dispersion and Gaussian-Hermite moments $h_3, h_4$)\footnote{The stellar kinematics of NGC 1023 and NGC 2685 were observed  and first presented in~\citet{2004MNRAS.352..721E}, and these observations were re-reduced as part of the Atlas$\rm ^{3D}$ Survey}.
The extended stellar kinematics was observed with the Mitchell integral field spectrograph~\citep{2008SPIE.7014E..70H} on the Harlan J. Smith telescope, as presented in~\citet{2017MNRAS.471.4005B}. The Mitchell integral field spectrograph has a large field-of-view of $1.68 \times 1.68$ arcmin$^2$, a large fibre size of $2.08$ arcsec and a maximum instrumental velocity resolution of $42$ km/s, and covers $2-4 R_\mathrm{e}$ for the sample galaxies. The stellar kinematics were obtained with the pPXF method which includes the upgrades described in~\citep{2017MNRAS.466..798C}, on spectra that were Voronoi binned to a target signal-to-noise ratio of 20. 

All of the sample galaxies were observed with the Westerbork Synthesis Radio Telescope (WSRT) and showed extended \HI discs or rings in the intensity maps~\citep{2012MNRAS.422.1835S}. However, only a subsample of four galaxies shows regular rotating features in their \HI velocity fields in deeper observations with the WSRT: NGC 2685, NGC 4203, NGC 5582 and NGC 6798 (see Appendix~\ref{sec:HI-fields} for these velocity maps). The observing time for these follow-up observations was $9 \times 12$ hours for each galaxy. The raw data were processed and calibrated using the \texttt{MIRIAD} software package~\citep{1995ASPC...77..433S}, resulting in a datacube with a spectral resolution of 16 km/s. We note that the observations of NGC 2685 and NGC 4203 were presented in~\citet{2009A&A...494..489J} and~\citet{2015MNRAS.451..103Y}, respectively, and refer the readers to~\citet{2015MNRAS.451..103Y} for more details about the data reduction.


\section{Methods}
\label{sec:method}
For the entire sample, we construct the Schwarzschild model with two-aperture stellar kinematics, the SAURON data and the Mitchell data. For the subsample with available WSRT data, we also build dynamical models combining two-aperture stellar kinematics and \HI velocity fields: the stellar kinematics is modelled with a Schwarzschild model, while the \HI velocity fields are modelled as a thin ring misaligned with the stellar disc. In this section, we use NGC 4203 from the subsample to demonstrate our methods. We begin by constructing the gravitational potential of the galaxy. We then illustrate the stellar orbit library used in the Schwarzschild models and the misaligned gas model for the \HI discs of the subsample. Subsequently, we select the best-fitting model and determine the $1-\sigma$ confidence levels. Finally, we obtain the mass budget, particularly the dark matter fraction, from the models.
 
\subsection{Gravitational potential}
\label{sec:gp}
The gravitational potential of the galaxy is contributed by three components: stars, a dark matter halo, and a central black hole. One key point to break the stellar-dark matter degeneracy is to describe the stellar mass distribution precisely. 

\begin{figure}
    \centering
    \includegraphics[width=0.8\columnwidth]{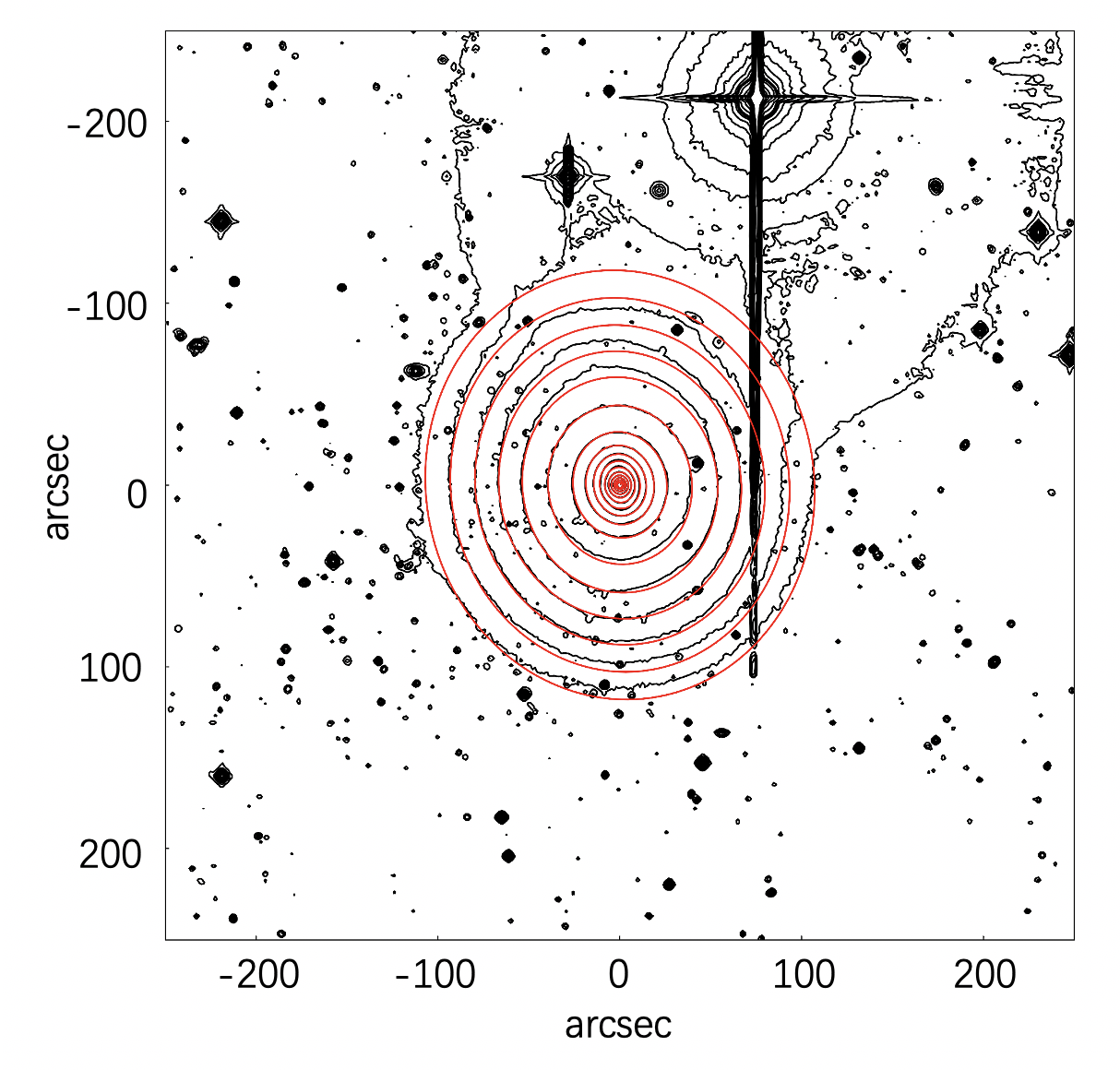}
    \includegraphics[width=0.8\columnwidth]{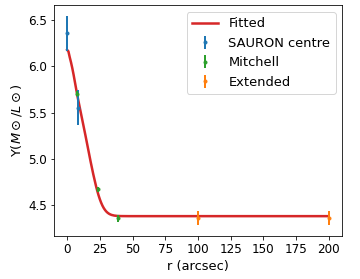}
    \caption{Top: MGE fitting of the $r$-band image of NGC 4203, with the surface brightness contours in black and its MGE model in red. The foreground star on the top right and its saturation were masked in the fitting. Bottom: $r$-band mass-to-light ratio gradient of NGC 4203. The blue dots and the corresponding uncertainties are the median values and the scatters of the $r$-band mass-to-light in each radial bin of~\citet{2017MNRAS.467.1397P}. The green dots and the corresponding uncertainties are the $V$-band mass-to-light ratio from~\citet{2017MNRAS.471.4005B} and linearly rescaled to match the $r$-band data. The orange dots are the extensions of the farthest green dot to 200 arcsec with twice the uncertainty to avoid nonphysical values or overfitting. The red curve is the best-fitting result.}
    \label{fig:ml-gradient}
\end{figure}

The stellar mass distribution of a galaxy is equal to its surface brightness times its stellar mass-to-light ratio. For the sample galaxies with no saturation in the central region of the MATLAS $r$-band image, we adopt the MATLAS $r$-band image as the surface brightness. For the rest of the sample galaxies, we first match and calibrate the Pan-STARRS images referring to the MATLAS $r$-band image; we then replace the saturated central region in the MATLAS $r$-band image with the Pan-STARRS one to generate a combined image; we finally adopt the combined image as the surface brightness, and mask foreground stars and their saturation. We fit the surface brightness with the sum of multiple 2-dimensional Gaussian profiles~\citep[MGE;][]{1994A&A...285..723E,2002MNRAS.333..400C}, and show the MGE fitting of NGC 4203 in the top panel of Figure~\ref{fig:ml-gradient}. 

The sample galaxies display strong gradients in their $r$-band mass-to-light ($M_*/L_r$) ratio obtained from the SAURON spectra~\citet{2017MNRAS.467.1397P} and the $V$-band mass-to-light ratio ($M_*/L_v$) obtained from the Mitchell spectra~\citep{2017MNRAS.471.4005B}, which are obtained with the stellar population synthesis models. Therefore, we take the mass-to-light ratio gradient into account instead of assuming a constant mass-to-light ratio~\citep{2020MNRAS.491.4221Y}. We first rescale the Mitchell $M_*/L_v$ linearly to match the SAURON $M_*/L_r$. We then generate a set of mass MGEs, ensuring that the corresponding mass-to-light ratio obtained from the mass and surface brightness MGEs is consistent with the observational one. We show the fitting of the mass-to-light gradient of NGC 4203 in the right panel of Figure~\ref{fig:ml-gradient}. 

The stellar mass and the corresponding mass-to-light ratio obtained through stellar population models are affected by the selection of the initial mass function (IMF)~\citep{2012A&A...538A..33S}. To account for this, we introduce a scaling factor $\alpha$ to indicate the total mass scale due to the IMF: $\alpha=1.0$ is set to the stellar mass produced by the Salpeter IMF~\citep{1955ApJ...121..161S}, while the Kroupa IMF~\citep{2001MNRAS.322..231K} produces a stellar mass roughly with $\alpha=0.6$~\citep{2014ApJS..214...15S}.

We then deproject the 2-dimensional MGE model to a triaxial MGE model with a set of viewing angles $(\theta,\psi,\phi)$. The viewing angles are linked to the three ratios of the intrinsic shape of the galaxy $(p,q,u)$~\citep{2008MNRAS.385..647V}, which are the free parameters of the gravitational potential. We set the minor-to-major axis ratio $q$ and the intermediate-to-major axis ratio $p$ as free parameters to allow triaxiality, and fix $u=0.9999$ to reduce model freedom. 

We choose a spherical generalised-NFW profile~\citep{1996ApJ...462..563N,1996MNRAS.278..488Z} for the dark matter halo, 
\begin{equation}
    \rho(r) = \frac{\rho_\mathrm{s}}{(r/r_\mathrm{s})^\gamma [1+(r/r_\mathrm{s})^\eta]^{(3-\gamma)/\eta}}.
\end{equation}
which contains four parameters: $\rho_s$ is the scale density, $r_s$ is the scale radius, $\gamma$ is the inner slope (when $\gamma = 1$, it reduces to the NFW profile), while $\eta$ controlling the turning point is fixed to $2$. 

The central black hole is regarded as a point source with a softening length. Because it is not resolved by the kinematic data and has little effect on the result, we fix its mass at $10^6 M_\odot$ and its softening length at $10^{-3}$ pc.

In total, we have six free parameters in the dynamical model: the total mass scale $\alpha$, the minor-to-major axis ratio $q$ and the intermediate-to-major axis ratio $p$, the scale density $\rho_s$, the scale radius $r_s$ and the inner slope $\gamma$ of the dark matter halo.

\subsection{Model of stellar kinematics}
We construct a Schwarzschild model for each set of parameters using the \texttt{DYNAMITE} package~\citep{2020ascl.soft11007J,2022A&A...667A..51T}, which is a modified version of a triaxial dynamical modelling code~\citep{2008MNRAS.385..647V} written in \texttt{Python}. We generate a general orbit library and an additional box library based on the orbit sampling method described in ~\citep{2020MNRAS.491.4221Y,2008MNRAS.385..647V}, and adopt a sampling number of each orbit library ($n_E \times n_\theta \times n_R$ for the general orbit library and $n_E \times n_\theta \times n_\phi$ for the box library) of $45 \times 15 \times 13$ with a dithering of $3$ in every dimension to fit the extended stellar kinematics.

The stellar kinematics have two apertures: the SAURON aperture has a higher resolution within a smaller field-of-view that is completely covered by the Mitchell one. To prevent the fitting from being biased toward the central region, we adjust the weights of each aperture in two steps: (1) we reduce the total weights of the SAURON aperture by a factor of $n_\mathrm{bins,S}/n_\mathrm{bins,M}$, where $n_\mathrm{bins,S}$ and $n_\mathrm{bins,M}$ are the number of bins for the SAURON and Mitchell apertures, respectively; (2) we increase the weights of the outer regions of the Mitchell aperture to make sure that their weights are comparable to the central region.

\subsection{Model of misaligned gas kinematics}
For each galaxy in the subsample, we measure the position angle ($\psi_1$) of the Mitchell stellar velocity field, and the position angle ($\psi_2$) and inclination ($i_2$) of the \HI velocity field using the package \texttt{KINEMETRY}~\citep{2006MNRAS.366..787K}, as shown in Table~\ref{tab:pa_incl}. There is a significant misalignment between the stellar disc (or the equatorial plane of the stellar component) and the \HI disc for NGC 4203, and a clear misalignment for NGC 2685 and NGC 6798 (about $15 \degree$ ignoring counter-rotation). The inclination of the stellar disc $i_1$ is actually the viewing angle $\theta$ introduced in Section~\ref{sec:gp}, which is bound to the stellar minor-to-major axis $q$. Considering that $q$ is a free parameter in the model, it is possible that there is a misalignment even for NGC 2685 when taking the misalignment in the inclinations into account. Therefore, we build a misaligned gas model to describe the motions of these \HI discs for all galaxies in the subsample. 
\begin{table}
    \centering
    \begin{tabular}{ccccc}
        \hline
        Galaxy & $\psi_1$ & $\psi_2$ & $\cos{i_2}$ & \\\hline
        NGC2685 & $53.5\degree$ & $40.5\degree$  & $0.267$ & \\
        NGC4203 & $-104.5\degree$ & $64.0\degree$ & $0.835$ & \\
        NGC5582 & $59.5\degree$ & $58.0\degree$ & $0.683$ & \\
        NGC6798 & $-49.0\degree$ & $145.0\degree$ & $0.525$ & \\\hline
    \end{tabular}
    \caption{Position angle (clock-wise from west to its major-axis) and inclination for the subsample. $\psi_1$: position angle measured from the Mitchell stellar velocity field; $\psi_2$ and $i_2$: position angle and inclination measured from the \HI velocity field, and we show $i_2$ in the form of $\cos{i_2}$ as it is the direct output of \texttt{KINEMETRY}.}
    \label{tab:pa_incl}
\end{table}

As the gas discs still show nearly regular rotation in their velocity fields, we assume that a particle on the gas disc circulates around its own short axis with a circular velocity $v_c$, and it reaches nearly dynamical equilibrium in the radial direction of the stellar disc. We project $v_c$ along the tangential direction of the stellar disc, 
\begin{equation}
    v_\phi = \frac{r_0 \cos{\delta}}{R_0} v_c, 
\end{equation}
where $r_0$ is its distance to the centre of the gas disc, $R_0$ is the distance to the short axis of the stellar disc, and $\delta$ is the angle between the axes of the stellar and gas disc. $\delta$ is connected to the position angle and inclination of the stellar disc $(\psi_1, i_1)$ and those of the gas disc $(\psi_2, i_2)$ by
\begin{equation}
\label{equ:misalign_angle}
    \cos \delta = \cos{i_1} \cos{i_2} \pm \sin{i_1} \sin{i_2} \cos{(\psi_1-\psi_2)},
\end{equation}
The $\pm$ sign takes into account the rotation direction of the inclination. In the most simplified case where two discs have the same position angle and inclination $(\psi, i)$, the $+$ sign represents the two discs fully aligned such that the model reduces to the \HI model described in \citet{2020MNRAS.491.4221Y}, and the $-$ sign represents the rare case where two discs rotate an angle of $i$ in opposite directions from the line of sight, so that $\delta = 2\times i$.

The equation for the dynamical equilibrium is 
\begin{equation}
    v_\phi = \sqrt{-\mathrm{AccR}_{R_0}\cdot R_0},
\end{equation}
where $\mathrm{AccR}_{R_0}$ is the gravitational acceleration along the radial direction of the stellar disc at the distance $R_0$ calculated from the gravitational potential. Given the mass distribution of a galaxy and the position on the gas disc, we can deduce the circular velocity $v_c$ of the gas disc. 

The modelled light-of-sight velocity of the gas disc is given by
\begin{equation}
v_\mathrm{mod} = v_c \sin{i_2} \cos{\phi'},
\end{equation}
where $\phi'$ is the azimuthal angle from the observed major axis of the gas disc. The modelled velocity is then convolved with the beam of the observation for a direct comparison with the observational velocity.

We will discuss this model of the misaligned gas kinematics further in Yang et al. in preparation.

\subsection{Confidence level}
We have a total of six free parameters. There are two parameters for the stellar component sampled on linear grids:
\begin{enumerate}
    \item $q$ is the minor-to-major axis ratio, $q \in [0.06,q_\mathrm{max}]$ with a step size of $0.02$, where $q_\mathrm{max}$ is the maximum intrinsic flattening allowed by the observed axis ratio of the surface brightness for each galaxy;
    \item $p$ is the intermediate-to-major axis ratio, $p \in [0.90,0.99]$ with a step size of $0.01$ for a nearly oblate intrinsic shape with some triaxiality still allowed. $p$ is further released to lower values in the case that the best-fitting value hits the boundary.
\end{enumerate}
There are three parameters for the dark matter profile:
\begin{enumerate}
    \item $\rho_\mathrm{s}$ is the central density sampled on a logarithmic grid, $\log[\rho_\mathrm{s}/(\mathrm{M_\odot}\cdot\mathrm{pc}^{-3})] \in [-5.00,-1.00]$ with a step of $0.25$; 
    \item $r_\mathrm{s}$ is the scale radius sampled on a logarithmic grid, $\log(r_\mathrm{s}/\mathrm{pc}) \in [3.75,5.50]$ with a step of $0.25$;
    \item $\gamma$ is the inner slope, $\gamma \in [0.0,2.0]$ on a linear grid with a step of $0.1$;
\end{enumerate}
The last is the total mass scale factor sampled on a linear grid:
\begin{enumerate}
    \item $\alpha$ represents the total mass produced by different IMFs, $\alpha \in [0.4,1.6]$ with a step of $0.1$.
\end{enumerate}

We explore the parameter grid by first creating the initial model from the given initial parameters and then computing the models adjacent to the initial model in the parameter grid. We select some models with the lowest $\chi^2$ and calculate new models next to them, repeating this step until no more new models are added. We finally choose the model with the least $\chi^2$ in the kinematic residual maps as the best-fitting model, with
\begin{equation}
\chi^2 = \chi_\mathrm{star}^2 (+ \chi_\mathrm{gas}^2).
\end{equation}

The fluctuation in the observational data results in a fluctuation in $\chi^2$. Considering that we include weights in the stellar kinematics, we adopt the bootstrapping method to measure this fluctuation for stellar kinematics. We add Gaussian perturbations whose standard deviations equal the observational $1-\sigma$ uncertainties to the observational stellar kinematics and measure its corresponding $\chi_\mathrm{bts}^2$. We repeat this process and take $\Delta \chi_\mathrm{star}^2 = \overline{\chi_\mathrm{bts}^2} + \sigma(\chi_\mathrm{bts}^2)-\chi_\mathrm{star,min}^2$ as the $1-\sigma$ confidence level for stellar kinematics, where $\overline{\chi_\mathrm{bts}^2}$ and $\sigma(\chi_\mathrm{bts}^2)$ are the mean and standard deviation of $\chi_\mathrm{bts}^2$ and $\chi_\mathrm{star,min}^2$ is the minimum of $\chi_\mathrm{star}^2$. 
We adopt $\Delta \chi_\mathrm{gas}^2 \sim \sqrt{4\chi_\mathrm{gas,min}^2 - 2N_\mathrm{gas}}$ as the $1-\sigma$ confidence level for gas kinematics (see Appendix~\ref{sec:clc}), where $\chi_\mathrm{gas,min}^2$ is the minimum of $\chi_\mathrm{gas}^2$ and $N_\mathrm{gas}$ is the number of gas data points. 
In the general case where the model is a simplification of the data, $\Delta \chi_\mathrm{gas}^2 \sim 2\sqrt{\chi_\mathrm{gas,min}^2}$, consistent with the bootstrapping result in~\citet{2020MNRAS.491.4221Y}. 
Therefore, the $1-\sigma$ confidence level is
\begin{equation}
\Delta \chi^2 =  \Delta \chi_\mathrm{star}^2 (+ \Delta \chi_\mathrm{gas}^2).
\end{equation}

\begin{figure*}
    \centering
    \includegraphics[width=\textwidth]{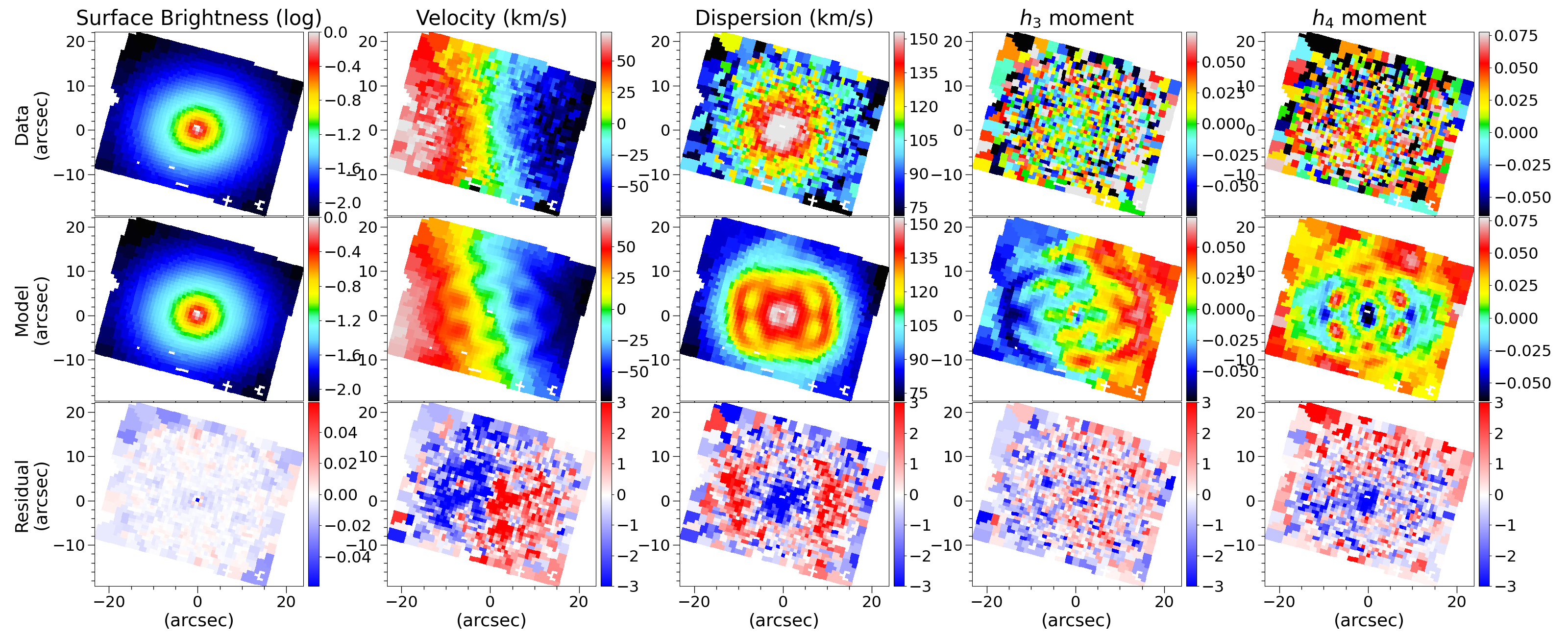}
    \includegraphics[width=\textwidth]{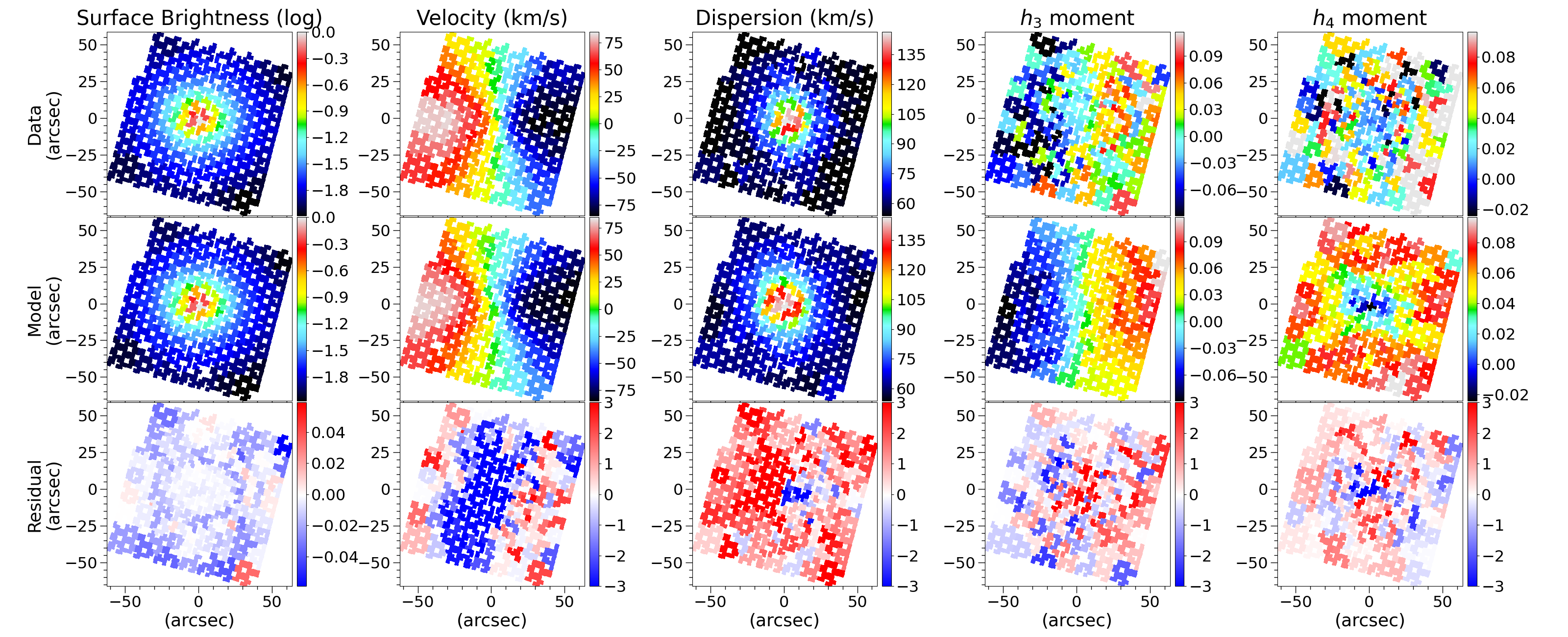}
    \includegraphics[width=\textwidth]{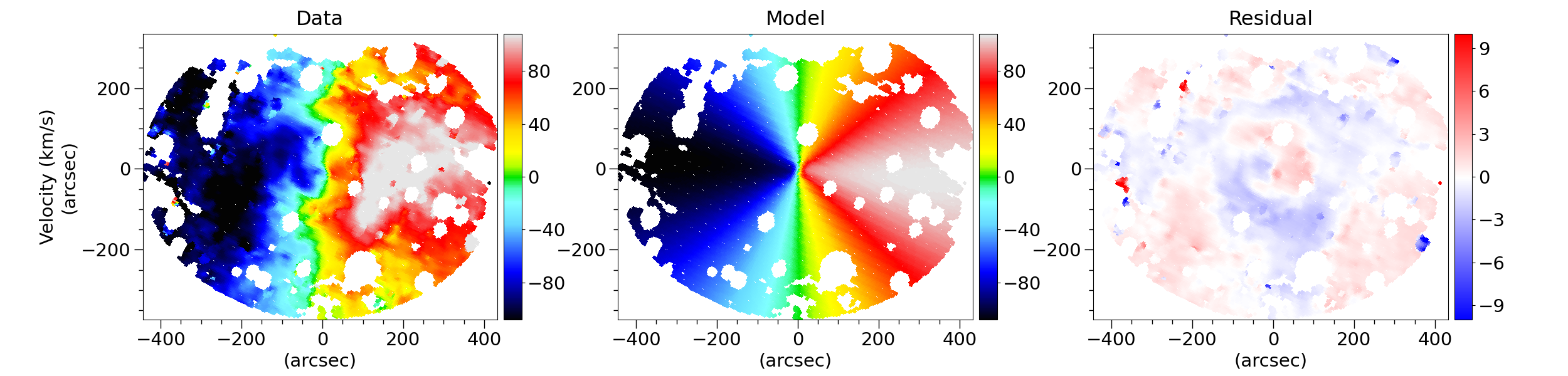}
    \caption{Best-fitting model with combined stellar and \HI kinematics of NGC 4203, with all maps rotated so that their major axis is aligned with the x-axis. The top and middle panels: SAURON and Mitchell stellar kinematics. In each of the top and middle panels: data (top), model (middle), and relative residual (bottom; defined as data-model/error) of the surface brightness, stellar velocity, velocity dispersion, and the third and fourth orders of Gauss–Hermite moments from left to right. The missing pixels in the Mitchell stellar are generated by gridding because of the gaps in the Mitchell fibres, which do not affect the dynamical modelling.
    The bottom panel: data (left), model (middle), and relative residual (right; defined as data-model/error) of the \HI velocity. We have masked the regions that are possibly the \HI blobs outside the disc, see Appendix~\ref{sec:HI-fields}.}
    \label{fig:kin}
\end{figure*}

We show the best-fitting model with the combined stellar and \HI kinematics of NGC 4203 in Figure~\ref{fig:kin} as an example of our fit, in which the major structures of the stellar kinematics and \HI velocity fields are recovered. 

\subsection{Mass budget}
\label{sec:mb}
We generate the enclosed mass profiles of the stellar, dark matter, and total mass from the model parameters for all sample galaxies to exhibit their mass distribution. For each galaxy, we first select all models within its $1-\sigma$ confidence level. We then obtain the enclosed mass profiles of stellar, dark matter and total mass for every model. We finally adopt the mean profile of all models within the $1-\sigma$ confidence level and take the standard deviation as the $1-\sigma$ uncertainties for the enclosed mass profiles of stellar, dark matter and total mass, respectively.

We obtain the dark matter fraction profiles, defined as the ratio of the enclosed dark matter and total mass within a certain radius, for all sample galaxies in a similar way. We first calculate the dark matter fraction profiles for every model within the $1-\sigma$ confidence level of each galaxy. We then adopt the mean dark matter fraction profile, and take the standard deviation as the $1-\sigma$ uncertainties.

\begin{figure}
    \centering
    \includegraphics[width=\columnwidth]{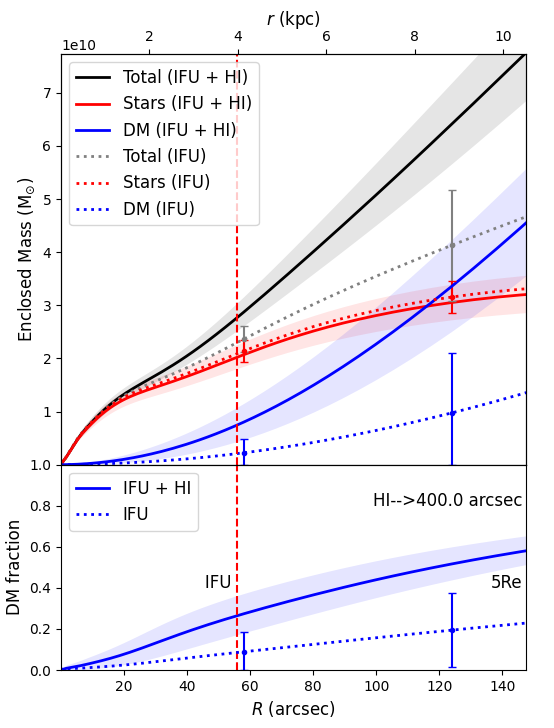}
    \caption{Top: Enclosed mass profile of NGC 4203. The black, red, and blue curves stand for the total, stellar, and dark matter mass obtained by our dynamical models; the solid curves are from the model with combined stellar and \HI kinematics, and the dotted curves are from the model with stellar kinematics. The corresponding shaded areas and typical errorbars show the $1-\sigma$ uncertainties for each curve.  Bottom: Dark matter fraction as a function of radius. The solid blue curve is from the model with combined stellar and \HI kinematics, and the dotted blue curve is from the model with stellar kinematics. The corresponding shaded area and typical errorbars show the $1-\sigma$ uncertainties for each line. The figure extends to $5 R_\mathrm{e}$, the red dashed lines mark the coverage of stellar kinematics. The \HI kinematics extend to $\sim 400$ arcsec as labelled in the figure.}
    \label{fig:enc}
\end{figure}

Figure~\ref{fig:enc} shows the enclosed mass profile and dark matter fraction profile of NGC 4203 from two sets of models: one with combined stellar and \HI kinematics and the other with stellar kinematics only. The figure extends to $5 R_\mathrm{e}$, and the coverage of stellar kinematics is marked with a red dashed line. The total enclosed mass profiles from the two models are consistent within the $1-\sigma$ uncertainties within the coverage of stellar kinematics: for NGC 4203, the model with stellar kinematics reports a total mass of $2.2_{-0.5}^{+0.6}\times10^{10} M_\odot$ while the model with combined stellar and \HI kinematics reports a total mass of $2.8_{-0.6}^{+0.6}\times10^{10} M_\odot$, and we also obtain similar consistency for the other galaxies in the subsample. However, the mass beyond is only accurately determined by the model with combined stellar and \HI kinematics. For NGC 4203, the stellar mass dominates the enclosed mass within $2 R_\mathrm{e}$ (approximately $5$kpc), while dark matter dominates outside of $5 R_\mathrm{e}$ (approximately $10$ kpc). The models constrained by stellar kinematics have considerable uncertainties in the dark matter mass, especially when extrapolating to the outer regions beyond the data coverage, even though both SAURON and Mitchell kinematics were used.

\section{Results}
\label{sec:result}
We list the best-fitting parameters using the two-aperture kinematics for all sample galaxies in Table~\ref{tab:best-fitting}. We also include the results of combining the stellar kinematics and the \HI misaligned discs for the subsample in additional columns for comparison and provide the orientation of the \HI discs calculated using Equation~\ref{equ:misalign_angle} in Table~\ref{tab:hi_discs}.

\begin{table*}
    \renewcommand\arraystretch{1.5}
    \centering
    \caption{Best-fitting parameters of the sample galaxies. $\alpha$ indicates the stellar mass scale factor; $q$ is the intrinsic flattening of the stellar components; $\rho_s$, $r_s$, and $\gamma$ are the central density, the scale radius, and the inner slope of the dark matter halo profile. The asterisks represent the parameters obtained with combined stellar and \HI misaligned discs for the sub-sample.}
    \begin{tabular}{c|c|c|c|c|c|c|c|c|c|c}
        \hline
        Galaxy & $\alpha$ & $\alpha^*$  & $q$ & $q^*$ & $\rho_s $ &  $\rho_s^*$ & $r_s$ & $r_s^*$ & $\gamma$ & $\gamma^*$ \\
         & & & & & \multicolumn{2}{c}{($10^{-3} {\rm M_\odot\,pc}^{-3}$)} & \multicolumn{2}{c}{(kpc)} & & \\ \hline
        NGC0680 & ${0.50} \pm {0.07}$ &  & ${0.66} \pm {0.08}$ &  & ${8.3}_{-6.4}^{+28.5}$ &  & ${30.4}_{-20.7}^{+64.5}$ &  & ${0.85} \pm {0.33}$ &  \\
        NGC1023 & ${0.60} \pm {0.03}$ &  & ${0.20} \pm {0.05}$ &  & ${1.6}_{-0.6}^{+1.1}$ &  & ${87.9}_{-42.7}^{+83.0}$ &  & ${0.97} \pm {0.15}$ &  \\
        NGC2685 & ${0.81} \pm {0.03}$ & ${0.81} \pm {0.03}$ & ${0.22} \pm {0.03}$ & ${0.23} \pm {0.03}$ & ${0.4}_{-0.2}^{+0.3}$ & ${0.4}_{-0.2}^{+0.3}$ & ${85.8}_{-42.3}^{+83.5}$ & ${78.1}_{-37.0}^{+70.4}$ & ${1.13} \pm {0.12}$ & ${1.14} \pm {0.12}$ \\
        NGC2764 & ${1.19} \pm {0.20}$ &  & ${0.10} \pm {0.05}$ &  & ${9.1}_{-4.4}^{+8.5}$ &  & ${12.1}_{-4.9}^{+8.2}$ &  & ${0.97} \pm {0.22}$ &  \\
        NGC3522 & ${0.59} \pm {0.11}$ &  & ${0.39} \pm {0.10}$ &  & ${3.4}_{-3.0}^{+27.8}$ &  & ${17.1}_{-14.4}^{+89.1}$ &  & ${0.94} \pm {0.29}$ &  \\
        NGC3998 & ${1.04} \pm {0.10}$ &  & ${0.31} \pm {0.08}$ &  & ${6.2}_{-4.9}^{+23.4}$ &  & ${9.9}_{-4.4}^{+7.8}$ &  & ${1.01} \pm {0.05}$ &  \\
        NGC4203 & ${0.47} \pm {0.05}$ & ${0.46} \pm {0.05}$ & ${0.63} \pm {0.10}$ & ${0.62} \pm {0.09}$ & ${0.8}_{-0.7}^{+4.3}$ & ${4.0}_{-2.8}^{+9.5}$ & ${14.9}_{-9.2}^{+23.9}$ & ${21.0}_{-11.8}^{+26.7}$ & ${0.92} \pm {0.44}$ & ${0.86} \pm {0.35}$ \\
        NGC5582 & ${0.78} \pm {0.06}$ & ${0.86} \pm {0.05}$ & ${0.47} \pm {0.06}$ & ${0.56} \pm {0.01}$ & ${1.5}_{-0.4}^{+0.6}$ & ${1.5}_{-0.1}^{+0.1}$ & ${25.4}_{-6.4}^{+8.6}$ & ${17.8}_{-7.8}^{+13.8}$ & ${1.36} \pm {0.12}$ & ${1.24} \pm {0.20}$ \\
        NGC5631 & ${0.72} \pm {0.12}$ &  & ${0.54} \pm {0.15}$ &  & ${11.7}_{-7.9}^{+24.8}$ &  & ${16.2}_{-10.0}^{+26.3}$ &  & ${0.76} \pm {0.45}$ &  \\
        NGC6798 & ${0.64} \pm {0.06}$ & ${0.90} \pm {0.05}$ & ${0.20} \pm {0.05}$ & ${0.34} \pm {0.02}$ & ${0.9}_{-0.5}^{+0.9}$ & ${0.8}_{-0.2}^{+0.2}$ & ${90.0}_{-40.2}^{+72.5}$ & ${17.8}_{-7.8}^{+13.8}$ & ${1.07} \pm {0.12}$ & ${1.24} \pm {0.15}$ \\
        UGC03960 & ${0.93} \pm {0.20}$ &  & ${0.67} \pm {0.09}$ &  & ${3.1}_{-2.0}^{+5.9}$ &  & ${30.8}_{-18.8}^{+48.2}$ &  & ${0.99} \pm {0.16}$ &  \\
        \hline
    \end{tabular}
    \label{tab:best-fitting}
\end{table*}

\begin{table}
    \centering
    \begin{tabular}{ccc}
        \hline
        Galaxy & Projection direction & $\delta$\\\hline
        NGC2685 & $-$ & $151.4\degree$\\
        NGC4203 & $-$ & $21.2\degree$\\
        NGC5582 & $-$ & $116.1\degree$\\
        NGC6798 & $+$ & $123.0\degree$\\\hline
    \end{tabular}
    \caption{Orientations of misaligned \HI discs. The projection direction represents $\pm$ in Equation~\ref{equ:misalign_angle}, and $\delta$ represents the angle between the rotating axes of the stellar and \HI discs.}
    \label{tab:hi_discs}
\end{table}

\subsection{Stellar mass}
The total stellar masses of the sample galaxies are affected by the choices of the IMFs used in the stellar population models, and we have introduced a scaling factor $\alpha$ to indicate this mass scale. $\alpha$ ranges from $\sim 0.5$ to $\sim 1.0$, suggesting a wide IMF distribution from Kroupa-like IMFs to Salpeter-like IMFs, and is not affected much by including the \HI kinematics. Previous research has found that there is an IMF variation in early-type galaxies, which correlates with factors such as the stellar mass-to-light ratio~\citep{2012Natur.484..485C} and velocity dispersion~\citep{2017ApJ...838...77L,2019MNRAS.485.5256Z}. However, we find no clear correlation between the preference of IMFs and other stellar or kinematic features within our sample, which is probably the result of a limited sample size.


Taking the scaling factor $\alpha$ into consideration, our full sample has a mean $r$-band stellar mass-to-light ratio of $3.5 M_\odot/L_\odot$ at the galaxy centre with a root mean square (rms) scatter of $1.2 M_\odot/L_\odot$, and $2.4 M_\odot/L_\odot$ at the galaxy outskirts with an rms scatter of $1.0 M_\odot/L_\odot$. 

\subsection{Dark matter fraction}
\label{sec:dm}
We obtain the dark matter fraction within $1R_\mathrm{e}$, $2R_\mathrm{e}$, and $5R_\mathrm{e}$ ($f_{\rm DM}(R_\mathrm{e})$, $f_{\rm DM}(2R_\mathrm{e})$, and $f_{\rm DM}(5R_\mathrm{e})$) from the dark matter fraction profiles derived in Section~\ref{sec:mb}, for all sample galaxies modelled with two-aperture stellar kinematics and for galaxies in the subsample modelled with combined stellar and \HI kinematics, as listed in Table~\ref{tab:dm-frac}. The full sample modelled with stellar kinematics has a mean dark matter fraction of $0.19\pm0.03$ with an rms scatter of $0.10$ within $1 R_\mathrm{e}$, $0.35\pm0.04$ with an rms scatter of $0.15$ within $2 R_\mathrm{e}$, and $0.62\pm0.05$ with an rms scatter of $0.18$ within $5 R_\mathrm{e}$. The subsample modelled with combined stellar and \HI kinematics has a mean dark matter fraction of $0.10\pm0.02$ with an rms scatter of $0.03$ within $1 R_\mathrm{e}$, $0.21\pm0.03$ with an rms scatter of $0.06$ within $2 R_\mathrm{e}$, and $0.47\pm0.03$ with an rms scatter of $0.13$ within $5 R_\mathrm{e}$, which is lower than the full sample on average but consistent with it within a $1-\sigma$ scatter.
\begin{table*}
    \renewcommand\arraystretch{1.5}
    \centering
    \caption{Dark matter fractions of all sample galaxies. The asterisks represent the dark matter fractions within $5 R_\mathrm{e}$ and $5$kpc obtained with combined stellar and \HI misaligned discs for the sub-sample.}
    \begin{tabular}{c|c|c|c|c|c|c|c|c}
        \hline
        Galaxy & $f_{\rm DM}(R_\mathrm{e})$ & $f_{\rm DM}(2R_\mathrm{e})$ & $f_{\rm DM}(5R_\mathrm{e})$ & $f_{\rm DM}^*(5R_\mathrm{e})$ & $f_{\rm DM}(1\mathrm{kpc})$ & $f_{\rm DM}(5\mathrm{kpc})$ & $f_{\rm DM}(10\mathrm{kpc})$ & $f_{\rm DM}^*(10\mathrm{kpc})$\\\hline
        NGC0680 & ${0.21}\pm{0.10}$ & ${0.44}\pm{0.13}$ & ${0.75}\pm{0.10}$ & - & ${0.06}\pm{0.04}$ & ${0.42}\pm{0.12}$ & ${0.67}\pm{0.12}$ & - \\
        NGC1023 & ${0.12}\pm{0.04}$ & ${0.29}\pm{0.06}$ & ${0.64}\pm{0.06}$ & - & ${0.04}\pm{0.02}$ & ${0.28}\pm{0.06}$ & ${0.53}\pm{0.07}$ & - \\
        NGC2685 & ${0.13}\pm{0.04}$ & ${0.28}\pm{0.06}$ & ${0.64}\pm{0.06}$ & ${0.62}\pm{0.06}$ & ${0.05}\pm{0.02}$ & ${0.34}\pm{0.06}$ & ${0.62}\pm{0.06}$ & ${0.60}\pm{0.07}$ \\
        NGC2764 & ${0.31}\pm{0.10}$ & ${0.48}\pm{0.10}$ & ${0.74}\pm{0.06}$ & - & ${0.22}\pm{0.11}$ & ${0.49}\pm{0.09}$ & ${0.70}\pm{0.07}$ & - \\
        NGC3522 & ${0.16}\pm{0.13}$ & ${0.30}\pm{0.19}$ & ${0.54}\pm{0.25}$ & - & ${0.12}\pm{0.11}$ & ${0.48}\pm{0.24}$ & ${0.65}\pm{0.25}$ & - \\
        NGC3998 & ${0.05}\pm{0.06}$ & ${0.12}\pm{0.12}$ & ${0.30}\pm{0.23}$ & - & ${0.03}\pm{0.04}$ & ${0.24}\pm{0.19}$ & ${0.40}\pm{0.27}$ & - \\
        NGC4203 & ${0.04}\pm{0.05}$ & ${0.09}\pm{0.10}$ & ${0.23}\pm{0.21}$ & ${0.58}\pm{0.07}$ & ${0.02}\pm{0.03}$ & ${0.11}\pm{0.11}$ & ${0.22}\pm{0.20}$ & ${0.56}\pm{0.07}$ \\
        NGC5582 & ${0.23}\pm{0.08}$ & ${0.39}\pm{0.10}$ & ${0.63}\pm{0.11}$ & ${0.40}\pm{0.08}$ & ${0.08}\pm{0.03}$ & ${0.29}\pm{0.09}$ & ${0.47}\pm{0.11}$ & ${0.26}\pm{0.08}$ \\
        NGC5631 & ${0.20}\pm{0.15}$ & ${0.39}\pm{0.19}$ & ${0.66}\pm{0.22}$ & - & ${0.07}\pm{0.09}$ & ${0.36}\pm{0.18}$ & ${0.58}\pm{0.22}$ & - \\
        NGC6798 & ${0.23}\pm{0.06}$ & ${0.45}\pm{0.08}$ & ${0.79}\pm{0.05}$ & ${0.29}\pm{0.06}$ & ${0.07}\pm{0.03}$ & ${0.38}\pm{0.08}$ & ${0.64}\pm{0.07}$ & ${0.19}\pm{0.05}$ \\
        UGC03960 & ${0.41}\pm{0.15}$ & ${0.64}\pm{0.13}$ & ${0.85}\pm{0.08}$ & - & ${0.16}\pm{0.10}$ & ${0.60}\pm{0.14}$ & ${0.78}\pm{0.09}$ & - \\
        \hline
    \end{tabular}
    \label{tab:dm-frac}
\end{table*}

\begin{figure*}
    \centering
    \includegraphics[width=\textwidth]{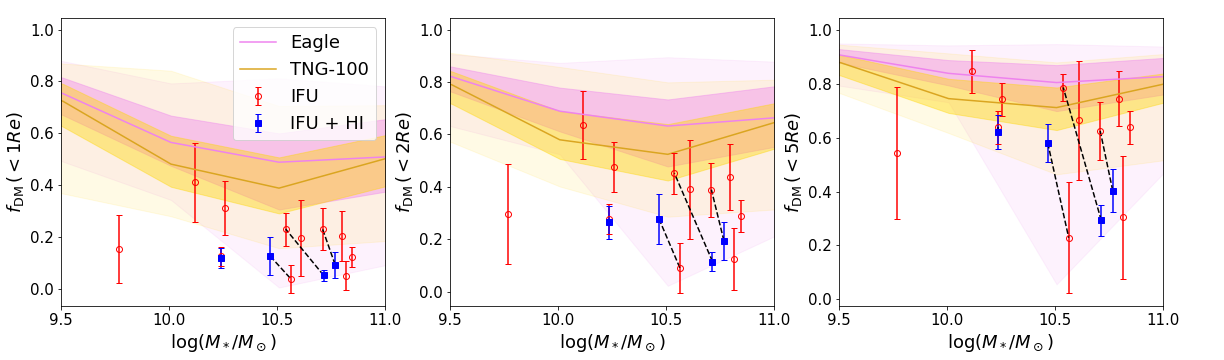}
    \includegraphics[width=\textwidth]{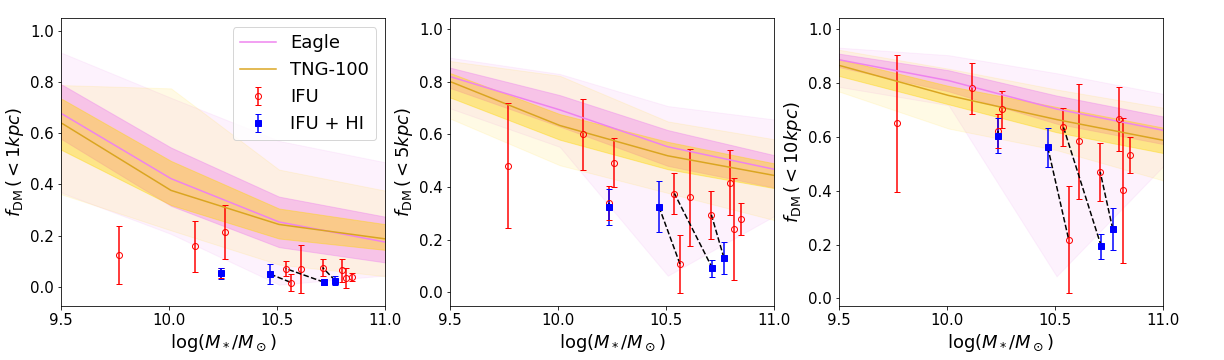}
    \caption{Relation of dark matter fraction and stellar mass for all sample galaxies. Top panel: from left to right are the dark matter fractions of the sample galaxies within $1 R_\mathrm{e}$, $2 R_\mathrm{e}$, and $5 R_\mathrm{e}$. Bottom panel: from left to right are the dark matter fractions of the sample galaxies within $1$kpc, $5$kpc, and $10$kpc. The empty red circles represent the results with two-aperture stellar kinematics, while the full blue squares represent the results with stellar and \HI kinematics, connecting to their counterparts in black dashed lines. For comparison, we plot the solid gold and violet curves (and the dark and light shadows) to represent the predictions (and the $1-\sigma$ and $3-\sigma$ uncertainties) of the TNG-100 simulation and the EAGLE simulation.}
    \label{fig:dmfrac-mass-my}
\end{figure*}

We plot the relation between the dark matter fractions and stellar mass for the sample galaxies in the top panel of Figure~\ref{fig:dmfrac-mass-my}. The results with two-aperture stellar kinematics are shown in empty red circles, whereas the results with stellar and \HI kinematics are shown in full blue squares, connecting to their counterpart results in black dashed lines. Among the four galaxies in the subsample, the dark matter fraction of NGC 2685 remains unchanged throughout all radii when include \HI kinematics, while the dark matter fraction of NGC 4203 increases, and the dark matter fractions of NGC 5582 and NGC 6798 decrease, especially at $f_{\rm DM}(5R_\mathrm{e})$. This is due to the coverage of stellar kinematics. As
Figure~\ref{fig:enc} shows, the total enclosed mass modelled with stellar kinematics is consistent with that modelled including \HI kinematics within the coverage of stellar kinematics, but the extrapolation to large radii is biased. The disparity in the fractions of stellar and dark matter further enlarges the bias in the dark matter fraction. Therefore, the dark matter fractions modelled with the stellar kinematics of a single galaxy can be randomly biased at large radii beyond the scope of the data, leading to unchanged, higher, or lower dark matter fractions. Nevertheless, the analysis of a large sample of galaxies can still provide some observational constraints on the dark matter fractions.

For comparison, we obtain analogous relations for the early-type galaxies in the TNG-100 and EAGLE simulations. We calculate its dark matter fractions, defined as the ratio of the mass of dark matter particles and that of all particles within a certain enclosed radius scaled by $R_\mathrm{e}$, using the same method as that published by the TNG-100 team~\citep{2018MNRAS.481.1950L}. We divide all simulated early-type galaxies within a mass range of $10^8-10^{12} M_\odot$ into mass bins with a logarithmic step of $0.5$ and take the median dark matter fraction of each mass bin as the value, the 16\% and 84\% fractiles as the $1-\sigma$ scatter, and the 0.15\% and 99.85\% fractiles as the $3-\sigma$ scatter. Galaxies in the EAGLE simulation have dark matter fractions systematically higher than the TNG-100 simulation, but with a much larger scatter towards low dark matter fractions.

There is a discrepancy between the observed dark matter fractions and the simulation predictions. Compared to the TNG-100 simulation, approximately 30\%, 40\%, and 70\% of the galaxies in the sample show $f_{\rm DM}(R_\mathrm{e})$, $f_{\rm DM}(2R_\mathrm{e})$, and $f_{\rm DM}(5R_\mathrm{e})$ that is consistent within a $1-\sigma$ scatter, while the remaining galaxies lie below the $1-\sigma$ region, and 20\% galaxies have dark matter fractions lying below the $3-\sigma$ region even at $5R_\mathrm{e}$. Compared to the EAGLE simulation, still 30\%, 40\% of the galaxies in the sample show $f_{\rm DM}(R_\mathrm{e})$ and $f_{\rm DM}(2R_\mathrm{e})$ that is consistent within a $1-\sigma$ scatter. However, only 40\% galaxies have $f_{\rm DM}(5R_\mathrm{e})$ which $1-\sigma$ is consistent with the EAGLE simulation, and almost no galaxy has $f_{\rm DM}(5R_\mathrm{e})$ lying below the $3-\sigma$ region.

Since the mass-size relation of this sample is not fully representative of the mass-size relations in the cosmological simulations, we make a similar plot using the dark matter fraction within $1$kpc, $5$kpc and $10$kpc ($f_{\rm DM}(1\mathrm{kpc})$, $f_{\rm DM}(5\mathrm{kpc})$ and $f_{\rm DM}(10\mathrm{kpc})$) to avoid the bias caused by $R_\mathrm{e}$ measurements, as shown in the bottom panel of Figure~\ref{fig:dmfrac-mass-my}. There is a turnover around $10^{10.5} M_\odot$ in the relations from the simulations when the radius is scaled by $R_\mathrm{e}$ in the top panel, but this turnover is not present when the absolute radii are used in the bottom panel, except for some galaxies with extremely low dark matter fractions around $10^{10.5} M_\odot$ in the EAGLE simulation.

The dark matter fractions observed using absolute enclosed radii also show similar discrepancies to the relations predicted by simulations, whose strength decreases with radius. The dark matter fractions within 1kpc, 5kpc, and 10kpc show similar discrepancies between observations and simulations as those within $1R_\mathrm{e}$, $2R_\mathrm{e}$, and $5R_\mathrm{e}$. This similarity suggests that the difference of $R_\mathrm{e}$ between the observed and simulated galaxies does not account for all the differences of enclosed dark-matter fractions between our sample galaxies and simulations.

\begin{figure*}
    \centering
    \includegraphics[width=\textwidth]{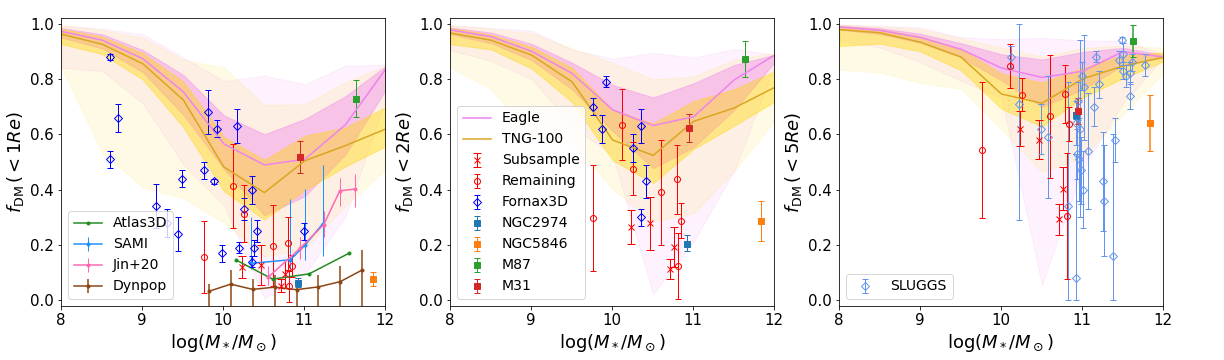}
    \caption{Relation of dark matter fraction and stellar mass from observations and simulations. From left to right: the dark matter fractions of the sample galaxies within $1 R_\mathrm{e}$, $2 R_\mathrm{e}$, and $5 R_\mathrm{e}$. Observations: this work (subsample using stellar IFU + \HI, remaining galaxies using stellar IFU), the galaxies in the Fornax3D project~\citep{2023arXiv230105532D}, the Atlas$\rm ^{3D}$ sample~\citep{2013MNRAS.432.1709C}, the SAMI sample~\citep{2022ApJ...930..153S}, the early-type galaxies presented in~\citet{2020MNRAS.491.1690J} from SDSS-IV MaNGA, early-type galaxies of the SDSS-IV MaNGA Dynpop sample~\citep{2024MNRAS.527..706Z}, the SLUGGS survey~\citep{2017MNRAS.468.3949A}, NGC2974~\citep{2020MNRAS.491.4221Y}, NGC5846~\citep{2016MNRAS.462.4001Z}, M87~\citep{2020MNRAS.492.2775L}, and M31~\citep{2022RAA....22h5023Y}. Simulations: TNG-100 simulation (in gold) and EAGLE simulation (in violet). }
    \label{fig:dmfrac-mass}
\end{figure*}

Dark matter fractions have been estimated using dynamical models with multiple tracers in previous studies. Stellar IFUs tracing stellar kinematics are typically used to measure dark matter fractions of galaxies within $1R_\mathrm{e}$, such as those of the Atlas$\rm ^{3D}$ sample~\citep{2013MNRAS.432.1709C}, the SAMI sample~\citep{2022ApJ...930..153S}, the early-type galaxies presented in~\citet{2020MNRAS.491.1690J} from the SDSS-IV MaNGA, early-type galaxies of the SDSS-IV MaNGA Dynpop sample~\citep{2024MNRAS.527..706Z} ,and the Fornax3D project~\citep{2023arXiv230105532D}. However, measurements can be extended to $5R_\mathrm{e}$ when combining stellar IFU with globular clusters (GCs), such as in the case of M 87~\citep{2020MNRAS.492.2775L}, or with GCs and planetary nebulae (PNe), such as in NGC 5846~\citep{2016MNRAS.462.4001Z} and M 31~\citep{2022RAA....22h5023Y}, or with \HI kinematics, such as in the case of NGC 2974~\citep{2020MNRAS.491.4221Y}. GCs alone can also be used to measure dark matter fractions beyond $5R_\mathrm{e}$, as seen in the SLUGGS survey~\citep{2017MNRAS.468.3949A}. We plot the correlation of dark matter fraction and stellar mass inferred from these measurements together with our results of this study, and compared it with analogous trends in the TNG-100 simulation and the EAGLE simulation, as shown in Figure~\ref{fig:dmfrac-mass}.

The observational dark matter fractions show a large scatter, and the $f_{\rm DM}(R_\mathrm{e})$ of~\citet{2023arXiv230105532D} and this work are higher than the median of the results obtained from the central stellar IFU data (Atlas$\rm ^{3D}$, SAMI and SDSS-IV MaNGA). All the observational data together still follow the same pattern as the simulations, with the dark matter fraction decreasing and then increasing as the stellar mass increases and reaching its minimum at a stellar mass of approximately $10^{10.5}M_\odot$. However, there is still an offset between the median dark matter fractions of the observational data and the relation of the simulations. This offset decreases as the enclosed radius increases: Various studies have measured $f_{\rm DM}(R_\mathrm{e})$, most of which have mean values below the $3-\sigma$ scatter of the relations predicted by the simulations. The sample of galaxies with $f_{\rm DM}(2-5\,R_\mathrm{e})$ in the literature is relatively small, and the galaxy fractions with dark matter fractions that are $1-\sigma$ consistent with simulation predictions are comparable to our sample. There are about 10\% of 47 galaxies with $f_{\rm DM}(5R_\mathrm{e})$ lying below the $3-\sigma$ region compared to the TNG-100 or the EAGLE simulations.



\subsection{Dark matter halo properties}
\label{sec:halo}
\begin{figure}
    \centering
    \includegraphics[width=\columnwidth]{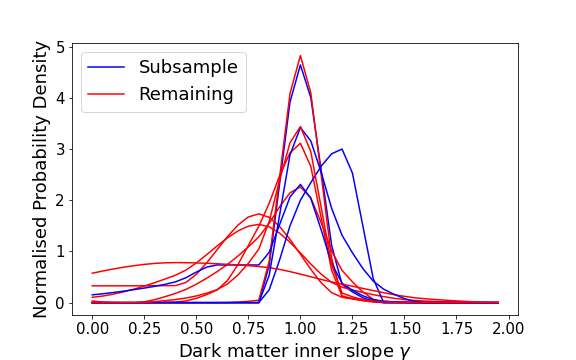}
    \caption{Normalised probability density distribution of dark matter inner slope $\gamma$ for all sample galaxies (subsample using stellar IFU + \HI, remaining galaxies using stellar IFU).}
    \label{fig:gamma-dist}
\end{figure}

The inner slopes of the dark matter profiles provide information to constrain the properties of dark matter. Our full sample modelled with stellar kinematics has uniform NFW-like dark matter cusps, which have a mean value of $1.00\pm0.04$ with an rms scatter of $0.15$. Including \HI kinematics does little change to the result: The subsample modelled with combined stellar and \HI kinematics has a mean value of $1.12\pm0.03$ with an rms scatter of $0.15$. 
Figure~\ref{fig:gamma-dist} displays the probability density distribution of the inner slopes of the dark matter profiles for our sample, which represents the normalised histogram of the inner slopes of all models within the $1-\sigma$ confidence level for each galaxy. Although the NFW-like dark matter cusps are favoured by our full sample modelled with stellar kinematics, there is still a possibility of dark matter cores for certain galaxies. The subsample modelled with combined stellar and \HI kinematics provides stronger constraints on the inner slopes of the dark matter profiles and largely eliminates other profiles except for the NFW-like cusps. Both models rule out steeper cusps with inner slopes greater than $1.5$.
Previous measurements indicate that density cusps ($\gamma \sim 1$) have been widely discovered in massive early-type galaxies~\citep[e.g.][]{2015ApJ...800...94S,2018ApJ...863..130W}, and also in dwarf spheroidals~\citep{2018MNRAS.481..860R,2020ApJ...904...45H}. Cuspy haloes ($\gamma \sim 0.5$) have also been reported in brightest cluster galaxies of relaxed galaxy clusters~\citep[e.g.][]{2004ApJ...604...88S,2013ApJ...765...25N} and nearby dwarf galaxies~\citep[e.g.][]{2014ApJ...789...63A,2022MNRAS.512.1012C}, but rarely reported in intermediate-mass early-type galaxies~\citep{2020MNRAS.491.4221Y}. Density cores usually exist in dwarf galaxies~\citep[e.g.][]{2003ApJ...596..957S,2008ApJ...681L..13B,2011ApJ...742...20W} and low surface brightness galaxies~\citep[e.g.][]{2001AJ....122.2396D,2008ApJ...676..920K}, but also in a few massive early-type galaxies~\citep{2010ApJ...716..370F,2018MNRAS.476..133O}.

\begin{figure}
    \centering
    \includegraphics[width=\columnwidth]{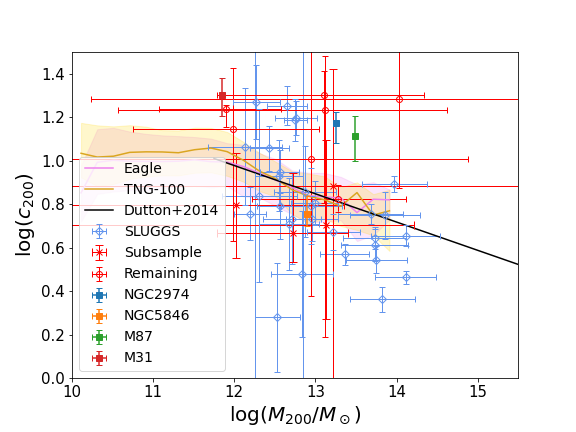}
    \caption{Relation of dark matter virial mass $M_{200}$ and concentration $C$ from observations and simulations. Observations: this work (subsample using stellar IFU + \HI, remaining galaxies using stellar IFU), the SLUGGS survey~\citep{2017MNRAS.468.3949A}, NGC2974~\citep{2020MNRAS.491.4221Y}, NGC5846~\citep{2016MNRAS.462.4001Z}, M87~\citep{2020MNRAS.492.2775L} and M31~\citep{2022RAA....22h5023Y}. Simulations: the TNG-100 simulation (in gold), the EAGLE simulation (in violet) and the relation presented in~\citet{2014MNRAS.441.3359D} (in black).}
    \label{fig:m200-c}
\end{figure}

The virial mass $M_{200}$ (defined as the enclosed mass within radius $r_{200}$, where the average density within $r_{200}$ is 200 times the critical density) and the concentration $C$ (defined as the ratio of $r_{200}$ and the scale radius $r_\mathrm{s}$) are commonly used to describe dark matter haloes, whose relation shows the evolution of dark matter haloes. We plot the observational relation between the virial mass $M_{200}$ and the concentration $C$ of dark matter haloes with that shown in the simulations, as shown in Figure~\ref{fig:m200-c}. The datasets are the same as those used in Figure~\ref{fig:dmfrac-mass}, except for the Fornax3D project which adopted the $M_{200}$-$C$ relation obtained with a simulation in Planck cosmology as presented in~\citet{2014MNRAS.441.3359D}. Therefore, we compare this relation with the $M_{200}$-$C$ relations in the TNG-100 and EAGLE simulations, as well as with the relation presented in~\citet{2014MNRAS.441.3359D}. The observational $M_{200}$-$C$ relation overlaps those in the simulations but shows a much larger scatter. 

\subsection{Total mass density}
Observations show that the total mass density of early-type galaxies follows an isothermal profile with a density slope of $\rho_\mathrm{tot} \propto r^{-2}$ and a small intrinsic scatter~\citep[e.g.][]{2015ApJ...804L..21C,2016MNRAS.460.1382S,2018MNRAS.476.4543B}, known as the `bulge-halo' conspiracy, which is also reproduced in cosmological simulations~\citep{2020MNRAS.491.5188W}. This isothermal profile is not a natural consequence in cold dark matter cosmology, and can be used to gain insight into the joint action of dark matter and stars in the galaxy formation process. Therefore, we obtain the total density profiles for our sample galaxies.

\begin{figure}
    \centering
    \includegraphics[width=\columnwidth]{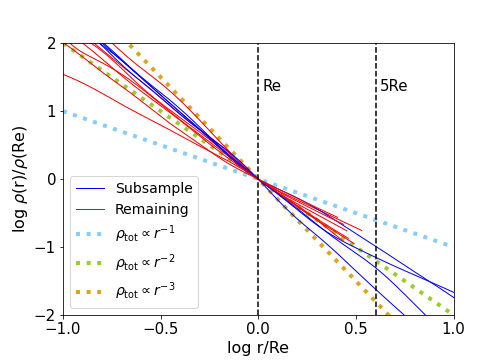}
    \caption{Profiles of total mass density distribution for all sample galaxies (subsample using stellar IFU + \HI, remaining galaxies using stellar IFU). $\rho \propto r^{-1}$ (NFW inner slope), $\rho \propto r^{-2}$ (isothermal profile) and $\rho \propto r^{-3}$ are shown in dashed lines for comparison.}
    \label{fig:rho_tot} 
\end{figure}

For each model within its $1-\sigma$ confidence level of a galaxy, we generate its total mass density profile from the derivative of its enclosed mass profile. We also fit the overall slope, the inner slope (between $0.1R_\mathrm{e}$ and $R_\mathrm{e}$) and the outer slope (between $R_\mathrm{e}$ and $4R_\mathrm{e}$) for each profile. We then adopt the mean profile of all models within the $1-\sigma$ confidence level and take their standard deviation as the $1-\sigma$ uncertainties for the total mass density profile. For the full sample modelled with stellar kinematics, we calculate the mean values and scatters of the overall, inner, and outer slopes using these slopes of all models within the $1-\sigma$ confidence level, respectively. The total density profiles of all sample galaxies follow a nearly isothermal profile that $\rho_\mathrm{tot} \propto r^{-2}$, and have a mean logarithmic slope $\gamma_\mathrm{tot} = 2.19\pm0.04$ with an rms scatter of $0.27$. There is a small difference between the inner and outer slopes: the mean inner slope is $2.27\pm0.04$ with an rms scatter of $0.31$, and the mean outer slope is $1.84\pm0.09$ with an rms scatter of $0.24$. Considering that only the galaxies in the subsample reach $4R_\mathrm{e}$, we calculate the average outer slope for the subsample modelled with combined stellar and \HI kinematics, $2.15\pm0.05$ with a scatter of $0.41$, finding no obvious systematic bias due to data coverage. We plot the distribution of the total mass density for sample galaxies in Figure~\ref{fig:rho_tot}.  

Our results are consistent with previous dynamical results using SAURON IFU and GC kinematics extended to $4R_\mathrm{e}$, which reports $\gamma_\mathrm{tot} = 2.19 \pm 0.03$ with an rms scatter $0.11$ in~\citet{2015ApJ...804L..21C} and $\gamma_\mathrm{tot} = 2.24 \pm 0.05$ in~\citet{2018MNRAS.476.4543B}, as well as dynamical results using \HI kinematics extended to $6R_\mathrm{e}$ in \citet{2016MNRAS.460.1382S}, which reports $\gamma_\mathrm{tot} = 2.18 \pm 0.03$ with an rms scatter $0.11$. 

\section{Discussion}
\label{sec:discussion}
\citet{2016MNRAS.460.3029B} modelled NGC 3998 with the same stellar kinematics data sets but a constant mass-to-light ratio and an NFW dark matter profile and reported $f_{\rm DM}(R_\mathrm{e}) = 0.07_{-0.07}^{0.08}$. In this work, we measure $f_{\rm DM}(R_\mathrm{e}) = {0.05}\pm 0.06$ for NGC 3998 taking into account the mass-to-light gradient, consistent with the previous result. Since the stellar mass dominates the enclosed mass within $1 R_\mathrm{e}$ in NGC 3998, the measurements of $f_{\rm DM}(R_\mathrm{e})$ are expected to be close to 0, regardless of the assumptions made.

The accuracy of dark matter fraction measurements depends strongly on the kinematic tracers available. For the full sample modelled with stellar kinematics, the dark matter fractions have relatively large uncertainties due to the low signal-to-noise ratio at the outskirts of the Mitchell stellar kinematic data, while combining \HI kinematic data can significantly reduce statistical errors for the subsample. There may be some systematic error caused by our misaligned gas model used to fit the \HI kinematics. This misaligned gas model is a first-order disc model in which the radial motions on the disc are not taken into account, which will be discussed further in the work of Yang et al. in preparation. The assumption that the gas disc is in nearly dynamical equilibrium is also a choice in the case of lacking boundary conditions in the \HI observations. Nevertheless, since this misaligned model reduces to the full aligned disc model when the angle between the stellar and \HI discs is reduced to zero, we believe that this assumption would have a minimal effect on the measurements of dark matter fractions. 

In Section~\ref{sec:dm}, we find that the dark matter fractions in observations are on average lower than the relations in the TNG-100 and the EAGLE simulations. Before concluding on a true discrepancy between observations and simulations, we would like to consider some impacts that may result in bias in the measurements of the dark matter fraction and then partially explain the discrepancy between the observational and simulated results.

\begin{enumerate}
\item Sample bias. The galaxies in the sample shown in Figure~\ref{fig:dmfrac-mass} are mainly early-type galaxies with intermediate masses, which may not reflect the overall characteristics of nearby early-type galaxies in terms of sizes, morphologies, star formation rate, etc. This could lead to a sample bias such that the results of the study may not be applicable to all nearby early-type galaxies.

\item $R_\mathrm{e}$ bias. Dark matter fractions are defined as the enclosed dark matter fraction within a certain radius, which is usually scaled with the half-light radii of the galaxy $R_\mathrm{e}$. However, galaxies in simulations are usually larger than those observed, particularly for early-type galaxies at the high-mass end~\citep{2018MNRAS.474.3976G,2021MNRAS.501.4359Z}. This bias in $R_\mathrm{e}$ may lead to a bias in the measurements of the dark matter fraction. In this work, we measured dark matter fractions on an absolute scale, as shown in the bottom panel of Figure~\ref{fig:dmfrac-mass-my} and found a similar discrepancy between the observed and simulated galaxies, suggesting that the bias in $R_\mathrm{e}$ is not a major factor in the measurements of the dark matter fraction.

\item Systematic bias of dynamical models. The errorbars provided in these dynamical measurements contain only statistical errors, while systematic errors are not included. Various assumptions used in dynamical models result in systematic bias: 
\begin{enumerate}
    \item The stellar mass-to-light ratios of galaxies tend to vary with the radius. Assuming a constant stellar mass-to-light ratio gradient instead of taking into account the actual gradient may lead to a significant underestimate of the dark matter fraction~\citep{2020MNRAS.492.2775L}.
    
    \item The choice of the formality of dark matter haloes may cause a systematic bias on the measurements of the dark matter fraction. ~\citep{2013MNRAS.432.1709C,2016MNRAS.462.4001Z}. One major contribution regards the inner slope of the dark matter profile: when the total enclosed mass of a galaxy is constrained at a certain radius, a density core with shallower inner slope will result in a lower dark matter fraction than a density cusp within the enclosed radius. Tests on mock core-like galaxies generated from simulation show that assuming an NFW dark matter halo, the dark matter fraction is overestimated by 38\%, while this overestimation reduces to 18\% if a generalised-NFW profile is used~\citep{2019MNRAS.486.4753J}. 
    
    \item The exact orbital anisotropies of tracers in galaxies are not known in advance. Assuming different orbital anisotropies can also lead to a systematic effect on the measurements of the dark matter fraction~\citep{2017MNRAS.468.3949A}.

\end{enumerate}
\end{enumerate}
We want to emphasise that our measurements have overcome most of these problems related to the systematic bias by considering stellar mass-to-light ratio gradient carefully, adopting a generalised-NFW profile for dark matter haloes, and creating a triaxial orbit-based Schwarzschild model. But dark matter fractions in the literature are measured with different assumptions and different data quality, which might lead to some systematic bias.

A fair comparison between simulations and observations could be achieved in the future by applying our method to a large sample of galaxies with uniformly high-quality data and ensuring that the galaxy samples from observations and simulations are representative of the nearby universe.

\section{Summary}
\label{sec:summary}
In this paper, we map the mass distributions of 11 early-type galaxies with two-aperture stellar kinematics that extended to $2-4 R_\mathrm{e}$ using the Schwarzschild model, and we map the mass distributions of a subsample of 4 galaxies with \HI discs using dynamical models that combine two-aperture stellar and \HI gas kinematics extended to $10 R_\mathrm{e}$. 
Our main findings are summarised below:
\begin{itemize}
    \item We adopt the stellar mass-to-light ratio obtained by stellar population synthesis assuming an Salpeter IMF, but allowing for a free scale parameter. The distribution of this scale parameter shows no preference for any particular IMF in our sample, suggesting a wide IMF distribution from Kroupa-like IMFs to Salpeter-like IMFs. 

    \item We robustly obtain the enclosed dark matter fraction within $\sim 5 R_\mathrm{e}$ for the 11 galaxies. The mean dark matter fraction within $1 R_\mathrm{e}$ is $0.2 \pm 0.1$ (rms), and increases to $0.4 \pm 0.2$ (rms) within $2 R_\mathrm{e}$, and $0.6 \pm 0.2$ (rms) within $5 R_\mathrm{e}$ for the full sample modelled with stellar kinematics, while for the subsample modelled with combined stellar and \HI kinematics, the dark matter fractions are $0.5 \pm 0.1$ (rms) within $5 R_\mathrm{e}$. The dark matter fractions of the sample galaxies are generally lower than the predictions of the TNG-100 and EAGLE simulations within $1 R_\mathrm{e}$, and there are 40\% and 70\% sample galaxies with dark matter fractions that are $1-\sigma$ consistent with either the TNG-100 or the EAGLE simulations at $2 R_\mathrm{e}$ and $5 R_\mathrm{e}$, respectively, while the remaining 60\% and 30\% galaxies lie below the $1-\sigma$ region. Combined with dark matter fraction measurements out to $5 R_\mathrm{e}$ in the literature, there are about 10\% of 47 galaxies lying below the $3-\sigma$ region of the TNG-100 or the EAGLE predictions.

    \item The sample galaxies show a preference for NFW-like dark matter haloes, with an average inner slope of $1.00\pm0.04$ with a root mean square (rms) scatter of $0.15$, and the results change little when including \HI kinematics. The dark matter virial mass $M_{\rm 200}$ and concentrated $C$ are only weakly constrained with large uncertainties, generally following the $M_{\rm 200}-C$ correlations from the TNG-100 and the EAGLE simulations, and the relation presented in~\citet{2014MNRAS.441.3359D}.
    
    \item The total mass density profiles are proportional to $r^{-2}$, consistent with the previous dynamical results with data coverage to $4 R_\mathrm{e}$~\citep{2015ApJ...804L..21C,2016MNRAS.460.1382S,2018MNRAS.476.4543B}. 
\end{itemize}

\section*{Acknowledgements}
The research presented here is supported by the CAS Project for Young Scientists in Basic Research, Grant No. YSBR-062. MY acknowledges the China Postdoctoral Science Foundation, Grant No. 2021M703337. NFB acknowledges the Science and Technologies Facilities Council (STFC) grant ST/V000861/1.

The WRST is operated by the ASTRON (Netherlands Institute for Radio Astronomy) with support from the Netherlands Foundation for Scientiﬁc Research (NWO).

The Pan-STARRS1 Surveys (PS1) and the PS1 public science archive have been made possible through contributions by the Institute for Astronomy, the University of Hawaii, the Pan-STARRS Project Office, the Max-Planck Society and its participating institutes, the Max Planck Institute for Astronomy, Heidelberg and the Max Planck Institute for Extraterrestrial Physics, Garching, The Johns Hopkins University, Durham University, the University of Edinburgh, the Queen's University Belfast, the Harvard-Smithsonian Center for Astrophysics, the Las Cumbres Observatory Global Telescope Network Incorporated, the National Central University of Taiwan, the Space Telescope Science Institute, the National Aeronautics and Space Administration under Grant No. NNX08AR22G issued through the Planetary Science Division of the NASA Science Mission Directorate, the National Science Foundation Grant No. AST-1238877, the University of Maryland, Eotvos Lorand University (ELTE), the Los Alamos National Laboratory, and the Gordon and Betty Moore Foundation.

\section*{Data Availability}
The data underlying in this article are available as follows: the MATLAS data at \url{http://matlas.astro.unistra.fr/WP/}, the PanSTARRS data at \url{https://outerspace.stsci.edu/display/PANSTARRS/}, the Atlas$\rm ^{3D}$ data at \url{http://www-astro.physics.ox.ac.uk/atlas3d/}, the Mitchell data at \url{https://doi.org/10.17630/43b4e7fb-2f2c-4867-a4d5-21599299c9d9}, the TNG-100 simulation at \url{https://www.tng-project.org/} and the EAGLE simulation at \url{https://eagle.strw.leidenuniv.nl/}. The \HI data underlying this article will be shared upon reasonable request to the corresponding authors.



\bibliographystyle{mnras}
\bibliography{reference} 




\appendix
\section{HI velocity fields}
\label{sec:HI-fields}
\begin{figure}
    \centering
    \includegraphics[width=0.9\columnwidth]{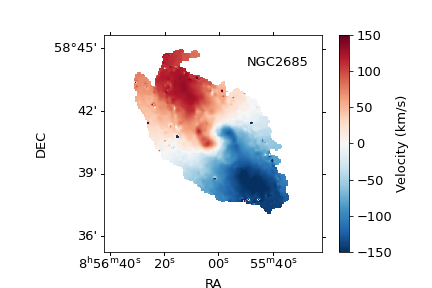}
    \includegraphics[width=0.9\columnwidth]{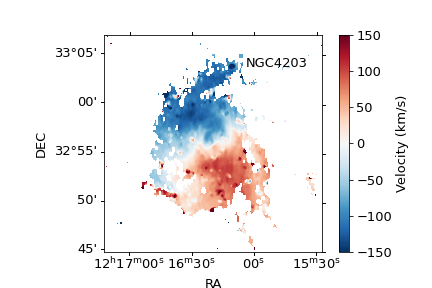}
    \includegraphics[width=0.9\columnwidth]{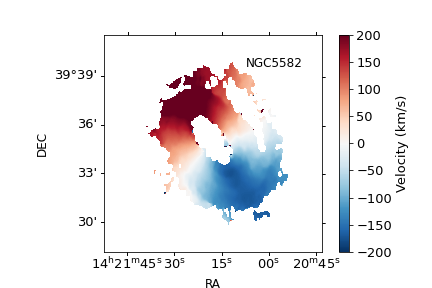}
    \includegraphics[width=0.9\columnwidth]{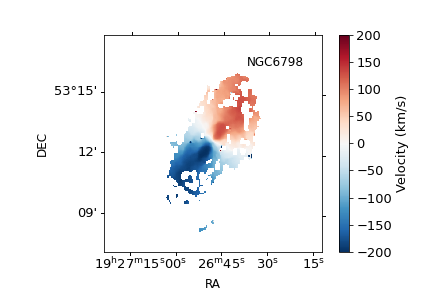}
    \caption{\HI velocity fields of the subsample obtained with the WSRT. The median velocity of each galaxy is already subtracted.}
    \label{fig:HImap-subsample}
\end{figure}

We exhibit the \HI velocity fields of the subsample obtained with WSRT in Figure~\ref{fig:HImap-subsample}.

An tilted inner ring is located in the centre of NGC 2685~\citep{2009A&A...494..489J}, so we mask the central region within an ellipse with a radius of $120$ arcsec.

There are some areas with discontinuous velocities in the velocity fields of NGC 4203, which are probably \HI blobs located outside the discs. Therefore, we mask these regions in our fitting of the misaligned gas models. 

There are extended arms in the velocity fields of NGC 4203 and NGC 5582, which are located outside $10 R_\mathrm{e}$ of each galaxy. As these arms might be biased from the nearly dynamical equilibrium, we mask the regions outside $10 R_\mathrm{e}$ for these galaxies.

\section{Confidence level calculation}
\label{sec:clc}
We first define an `ideal' residual of the best-fitting model by
\begin{equation}
\chi_\mathrm{ideal}^2 = \sum_i^N \left(\frac{\mu_{\mathrm{mod},i}-\mu_{\mathrm{data},i}}{\epsilon_{\mathrm{data},i}}\right)^2,
\end{equation}
where $f_{\mathrm{mod},i}$ is the best-fitting modelled data, $\mu_{\mathrm{data},i}$ and $\epsilon_{\mathrm{data},i}$ is the `real' data and its uncertainty, and $N$ is the number of independent data points. We assume the observation uncertainty $\epsilon_{\mathrm{obs},i} = \epsilon_{\mathrm{data},i}$, then possible observational data follow a Gaussian distribution that $f_{\mathrm{obs},i} \sim \mathcal{N}(\mu_{\mathrm{data},i},\,{\epsilon_{\mathrm{obs},i}}^2)$.

The residual distribution for the possible observation data is obtained by
\begin{align}
\label{mitequ:chi2}
\begin{split}
        \chi_\mathrm{f}^2 &= \sum_i^N \left(\frac{\mu_{\mathrm{mod},i}-f_{\mathrm{obs},i}}{\epsilon_{\mathrm{obs},i}}\right)^2 \\
        &= \chi_\mathrm{ideal}^2 +\sum_i^N {x_i^2} - \sum_i 2\left(\frac{f_{\mathrm{mod},i}-\mu_{\mathrm{data},i}}{\epsilon_{\mathrm{obs},i}}\right) x_i,
\end{split}
\end{align}
where each $x_i \sim \mathcal{N}(0,\,1)$. Its expectation $\mathbb{E}(\chi_\mathrm{f}^2) = \chi_\mathrm{ideal}^2 + N$, and its standard deviation $\sigma(\chi_\mathrm{f}^2) = \sqrt{4\chi_\mathrm{ideal}^2+2N} = \sqrt{4\,\mathbb{E}(\chi_\mathrm{f}^2) - 2N}$.

For the given observational data, we expect its residual $\chi_\mathrm{obs}^2 \sim \mathbb{E}(\chi_\mathrm{f}^2)$. Therefore, we adopt $\Delta \chi_\mathrm{obs}^2 = \sigma(\chi_\mathrm{f}^2) \sim \sqrt{4\chi_\mathrm{obs}^2 - 2N}$ as the $1-\sigma$ confidence interval due to observation uncertainties.




\bsp	
\label{lastpage}
\end{document}